\documentclass[a4paper,epsfig]{jpconf}
\usepackage{graphicx}
\def \lleq {\lower0.9ex\hbox{ $\buildrel < \over \sim$} ~}
\def \ggeq {\lower0.9ex\hbox{ $\buildrel > \over \sim$} ~}

\def \ol   {\Omega_{\Lambda}}

\def \omm  {\Omega_{0 {\rm m}}}

\def \beq  {\begin{equation}}
\def \eeq  {\end{equation}}
\def \ber  {\begin{eqnarray}}
\def \eer  {\end{eqnarray}}

\newcommand{\newc}{\newcommand}
\newc{\be}{\begin{equation}}
\newc{\ee}{\end{equation}}
\newc{\ba}{\begin{eqnarray}}
\newc{\ea}{\end{eqnarray}}
\newc{\bea}{\begin{eqnarray*}}
\newc{\eea}{\end{eqnarray*}}
\newc{\D}{\partial}
\newc{\ie}{{\it i.e.} }
\newc{\eg}{{\it e.g.} }
\newc{\etc}{{\it etc.} }
\newc{\lcdm }{$\Lambda$CDM }

\newc{\ra}{\rightarrow}
\newc{\lra}{\leftrightarrow}
\newc{\gsim}{\buildrel{>}\over{\sim}}
\newcommand {\ga} {\ {\raise-.5ex\hbox{$\buildrel>\over\sim$}}\ }
\newcommand {\la} {\ {\raise-.5ex\hbox{$\buildrel<\over\sim$}}\ }

\renewcommand{\[}{\left[}
\renewcommand{\]}{\right]}
\begin{document}
\title{$\Lambda$CDM: Triumphs, Puzzles and Remedies}

\author{L. Perivolaropoulos}

\address{Department of Physics, University of Ioannina, Greece}

\ead{leandros@uoi.gr}
\begin{abstract}
The consistency level of \lcdm with geometrical data probes has been increasing with time during the last decade. Despite of these successes, there are some puzzling conflicts between \lcdm predictions and dynamical data probes  (bulk flows, alignment and magnitude of low CMB multipoles, alignment of quasar optical polarization vectors, cluster halo profiles).  Most of these puzzles are related to the existence of  preferred anisotropy axes  which appear to be unlikely close to each other. A few models that predict the existence of preferred cosmological axes are briefly discussed.
\end{abstract}



\vspace{5mm}
A wide range of precise cosmological observations (Hicken {\it et al.}, 2009; Astier {\it et al.}, 2006; Kowalski {\it et al.}, 2008; Komatsu {\it et al.}, 2009; Reid {\it et al.}, 2010) that developed during the past two decades are well described by a class of cosmological models that rely on a set of simple assumptions:
\begin{itemize}
\item
The universe is homogeneous and isotropic on scales larger than a few hundred Mpc.
\item
General Relativity is the correct theory that describes gravity on all macroscopic scales.
\item
The universe consists of radiation (photons), matter (dark matter, baryons and leptons) and dark energy (a substance with repulsive gravitational properties which dominates at recent cosmological times and leads to accelerating cosmic expansion (Copeland, Sami \& Tsujikawa, 2006)).
\item
Primordial fluctuations that gave rise to structure formation were created as quantum fluctuations in an approximately scale invariant process that took place during inflation.
\end{itemize}
The simplest representative of the above class of models is the \lcdm model (Sahni, 2002; Padmanabhan, 2003). In this model the role of dark energy is played by the {\it cosmological constant}, a homogeneous form of energy whose density remains constant even in an expanding background. This is the current {\it standard cosmological model} and it is consistent with the vast majority of cosmological observations. Such observations involve geometric probes (Hicken {\it et al.}, 2009; Astier {\it et al.} 2006; Kowalski {\it et al.}, 2008; Komatsu {\it et al.}, 2009; Reid {\it et al.}, 2010) (direct probes of the large scale cosmic metric) and dynamical probes (Bertschinger, 2006; Nesseris \& Perivolaropoulos, 2008) of the large scale cosmic structure that probe simultaneously the large scale cosmic metric and the gravitational growth of perturbations, namely the theory of gravity on large scales.

Geometric probes of the cosmic expansion include the following:
\begin{itemize}
\item Type Ia supernovae (SnIa) standard candles (Hicken {\it et al.}, 2009; Astier {\it et al.}, 2006; Kowalski {\it et al.}, 2008).
\item The angular location of the first peak in the CMB perturbations angular power spectrum(Komatsu {\it et al.}, 2009). This peak probes the integrated cosmic expansion rate using the last scattering horizon as a standard ruler.
\item Baryon acoustic oscillations of the matter density power spectrum. These oscillations also probe the integrated cosmic expansion rate on more recent redshifts using the last scattering horizon as a standard ruler (Reid {\it et al.}, 2010).
\item Other less accurate standard candles (Gamma Ray Bursts (Basilakos  \& Perivolaropoulos, 2008), HII starburst galaxies (Plionis {\it et. al.}, 2009)) and standard rulers (cluster gas mass fraction (Allen {\it et. al.}, 2004) as well as probes of the age of the universe (Krauss \& Chaboyer, 2003).
\end{itemize}
Dynamical probes of the cosmic expansion and the gravitational law on cosmological scales include:
\begin{itemize}
\item X-Ray cluster growth data (Rapetti {\it et. al.}, 2008).
\item Power spectrum of Ly-$\alpha$ forest at various redshift slices  (McDonald {\it et al.}, 2005; Nesseris \& Perivolaropoulos, 2008).
\item Redshift distortion observed through the anisotropic pattern of galactic redshifts on cluster scales (Hawkins {\it et al.}, 2003)
\item Weak lensing surveys (Benjamin {\it et al.}, 2007; Amendola, Kunz \& Sapone, 2008)
\end{itemize}
These cosmological observations converge on the fact that the simplest model describing well the cosmic expansion rate is the one corresponding to a cosmological constant (Padmanabhan, 2003) in a flat space namely
\be H(z)^2=H_0^2 \[\omm (1+z)^3 + \ol \] \label{hzlcdm} \ee where $H(z)$ is the Hubble expansion rate at redshift $z$, $H_0=H(z=0)$, $\omm$ the present matter density normalized to the present critical density for flatness and $\ol = 1-\omm$ is the normalized dark energy density which is time independent in the simplest case of the cosmological constant (\lcdm).

In view of the wide range of successful predictions of \lcdm, three possible approaches develop for cosmological research:
\begin{itemize}
\item
{\bf Mainstream Observers Approach:} Supporters of this approach focus on the majority of cosmological data that are consistent with \lcdm. Thus, one assumes validity of \lcdm and uses cosmological observations to impose constraints on the model parameters (such as $\omm$) with the best possible accuracy. The advantage of this approach is that given the present status of cosmological observations, it is the most likely to lead to accurate physical results. On the other hand, this approach is unlikely to reveal any new physics beyond \lcdm if such physics is hidden in the data.
\item
{\bf Theorist's Approach:} This approach focuses on theoretical motivation and uses intuition and theoretical appeal to construct models more general than \lcdm which usually include the standard model as a special point in parameter space. In this approach, the parameter space of the theory is initially enlarged in directions motivated by theoretical arguments. Subsequently, cosmological observations are used to constrain this parameter space in a region which is usually around the point corresponding to \lcdm. The advantage of this approach is that it can produce beautiful and exciting theoretical results and predictions. On the other hand, it is unlikely to lead to the discovery of new physics because the simplicity of \lcdm makes it a preferable model -in the context of a Bayesian approach- compared to any more complicated theoretical model.
\item
{\bf Outlier Data Approach:} This approach focuses on the minority of data (outliers) that are inconsistent with \lcdm at a level of more than $2-3\sigma$. Then one identifies common features of these data and constructs theoretical models consistent with these features. These models are used to make non-trivial predictions for upcoming cosmological observations. The construction of these models is not guided by theoretical motivation but by existing data which however may be affected by systematic or large statistical fluctuations. The disadvantage of this approach is that there is a relatively high probability that these `outlier' data may be infected by large systematic or statistical fluctuations. As a result, the corresponding theoretical models may turn out to be unrealistic by future observations. On the other hand, if the `outlier' data turn out to be representative of the real world, this approach is the most likely to reveal the existence of new physics. Historically, it may be verified that indeed this approach has led to the discovery of new models that constitute better descriptions of Nature than previous `standard models'. For example, in the early '90s preliminary `outlier' data (Efstathiou, Sutherland \& Maddox, 1990) had challenged the sCDM model (flat $\omm=1$) which was at the time the `standard' cosmological model. Such data had provided early hints that $\omm<1$ but at the time they were considered systematic or statistical fluctuations. Only after the SnIa data (Perlmutter {\it et al.}, 1999), it was realized that the sCDM model needs to be abandoned in favor of \lcdm.
\end{itemize}
Therefore, the question that needs to be addressed is the following: {\it Are there currently similar data that challenge the current standard model (\lcdm) and what are their common features?} The answer to this question is positive. Indeed, these challenging to \lcdm data may be summarized as follows (Perivolaropoulos, 2008):
\begin{enumerate}
\item
{\bf Large Scale Velocity Flows:} \lcdm predicts significantly smaller amplitude and scale of flows than what observations indicate. It has been found that the dipole moment (bulk flow) of a combined  peculiar velocity sample extends on scales up to $100 h^{-1}Mpc$ ($z\leq 0.03$) with amplitude larger than $400 km/sec$ (Watkins, Feldman \& Hudson 2009). The direction of the flow has been found consistently to be approximately in the direction $l \simeq 282^\circ$, $b\simeq 6^\circ$. Other independent studies have also found large bulk velocity flows on similar directions on scales of about $100 h^{-1}Mpc$ (Lavaux {\it et. al.}, 2010) or larger (Kashlinsky {\it et. al.}, 2009). The expected $rms$ bulk flow in the context of \lcdm normalized with WMAP5 $(\omm,\sigma_8)=(0.258,0.796)$ on scales larger than $50h^{-1}Mpc$ is approximately $110 km/sec$. The probability that a flow of magnitude larger than $400km/sec$ is realized in the context of the above \lcdm normalization (on scales larger than $50h^{-1} Mpc$) is less than $1\%$. A possible connection of such large scale velocity flows and cosmic acceleration is discussed by Tsagas (2010).
\item
{\bf Alignment of low multipoles in the CMB angular power spectrum:} The normals to the octopole and quadrupole planes are aligned with
the direction of the cosmological dipole at a level inconsistent with Gaussian random, statistically isotropic skies at 99.7\% (Copi {\it et. al.}, 2010). The corresponding directions are: octopole plane normal $(l,b)=(308^\circ, 63^\circ)$, quadrupole plane normal $(l,b)=(240^\circ, 63^\circ)$ (Tegmark, de Oliveira-Costa \& Hamilton 2003), CMB dipole moment $(l,b)=(264^\circ, 48^\circ)$ (Lineweaver {\it et. al.}, 1996). A related effect has also been recently observed by considering the temperature profile of 'rings' in the WMAP temperature fluctuation maps (Kovetz,  Ben-David  \& Itzhaki, 2010). It was found that there is a ring with anomalously low mean temperature fluctuation with axis in the direction $(l,b)=(276^\circ, -1^\circ)$ which is relatively close to the above directions (particularly that corresponding to the bulk velocity flows).
\item
{\bf Large scale alignment in the QSO optical polarization data:} Quasar polarization vectors are not randomly oriented over the sky with a probability often in excess of 99.9\%. The alignment effect seems to be prominent along a particular axis in the direction $(l,b)=(267^\circ, 69^\circ)$ (Hutsemekers {\it et. al.}, 2005).
\item
{\bf Profiles of Cluster Haloes:} \lcdm predicts shallow low concentration and density profiles in contrast to observations which indicate denser high concentration cluster haloes (Broadhurst {\it et. al.}, 2005; Umetsu \& Broadhurst, 2008).
\item
{\bf Missing power on the low $l$ multipoles of the CMB angular power spectrum} which leads to a vanishing correlation function $C(\theta)$ on angular scales larger than $60^\circ$ (Copi {\it et. al.}  2007; Copi  {\it et. al.}, 2010)
\end{enumerate}
In addition to the above large scale effects there are issues on galactic scales (missing satellites
problem (Klypin {\it et. al.}, 1999)  and the cusp/core nature
of the central density profiles of dwarf galaxies (Gentile {\it et. al.}, 2004)).

Three of the above five large scale puzzles are large scale effects related to preferred cosmological directions (CMB multipole alignments, QSO polarization alignment and large scale bulk flows) which appear to be not far from each other (Antoniou \& Perivolaropoulos, 2010). Their direction is approximately normal to the axis of the ecliptic poles $(l,b)=(96^\circ,30^\circ)$ and lies close to the ecliptic plane and the equinoxes. This coincidence has triggered investigations for possible systematic effects related to the CMB preferred axis but no significant such effects have been found (Copi {\it et. al.}, 2010).

In addition, it has been shown recently (Antoniou \& Perivolaropoulos, 2010) that the Union2 SnIa data hint towards a direction of maximum accelerating expansion that is abnormally close to the directions of the above preferred axes. In Table 1, I show the directions of the preferred axes from different cosmological observations along with the corresponding references.
\vspace{0pt}
\begin{table}[!t]
\begin{center}
\caption{Directions of Preferred axes from different cosmological observations.}
\begin{tabular}{cccc}
\hline
\hline\\
\vspace{3pt}\textbf{Cosmological Obs.} \hspace{0pt}\& \textbf{\it l}\hspace{3pt} & \textbf{\it b}\hspace{3pt}
\&\textbf{Reference}\hspace{1pt}  \\
\vspace{3pt} SnIa Union2                  & $309^\circ$               \hspace{7pt} &  $18^\circ$      \hspace{7pt} &  (Antoniou \& Perivolaropoulos, 2010)             \\
\vspace{3pt} CMB Dipole                     & $264^\circ$                \hspace{7pt} &  $48^\circ$       \hspace{7pt} &   (Lineweaver {\it et. al.}, 1996)          \\
\vspace{3pt} Velocity Flows                  & $282^\circ$               \hspace{7pt} &  $6^\circ$      \hspace{7pt} &  (Watkins, Feldman \& Hudson 2009)                  \\
\vspace{3pt} Quasar Alignment           & $267^\circ$                \hspace{7pt} &  $69^\circ$   \hspace{7pt} &  (Hutsemekers {\it et. al.}, 2005)                    \\
\vspace{3pt} CMB Octopole             & $308^\circ$                \hspace{7pt} &  $63^\circ$   \hspace{7pt} &  (Bielewicz, Gorski \&  Banday, 2004)                  \\
\vspace{3pt} CMB Quadrupole                & $240^\circ$               \hspace{7pt} &  $63^\circ$      \hspace{7pt} &  (Bielewicz, Gorski \&  Banday, 2004)                   \\
\hline \\
\vspace{3pt} Mean               & $278^\circ \pm 26^\circ$               \hspace{7pt} & $45^\circ \pm 27^\circ$    \hspace{7pt} &           -                  \\
\hline \hline
\end{tabular}
\end{center}
\end{table}
\begin{figure}[!b]
\begin{center}
\includegraphics[width=3.5in]{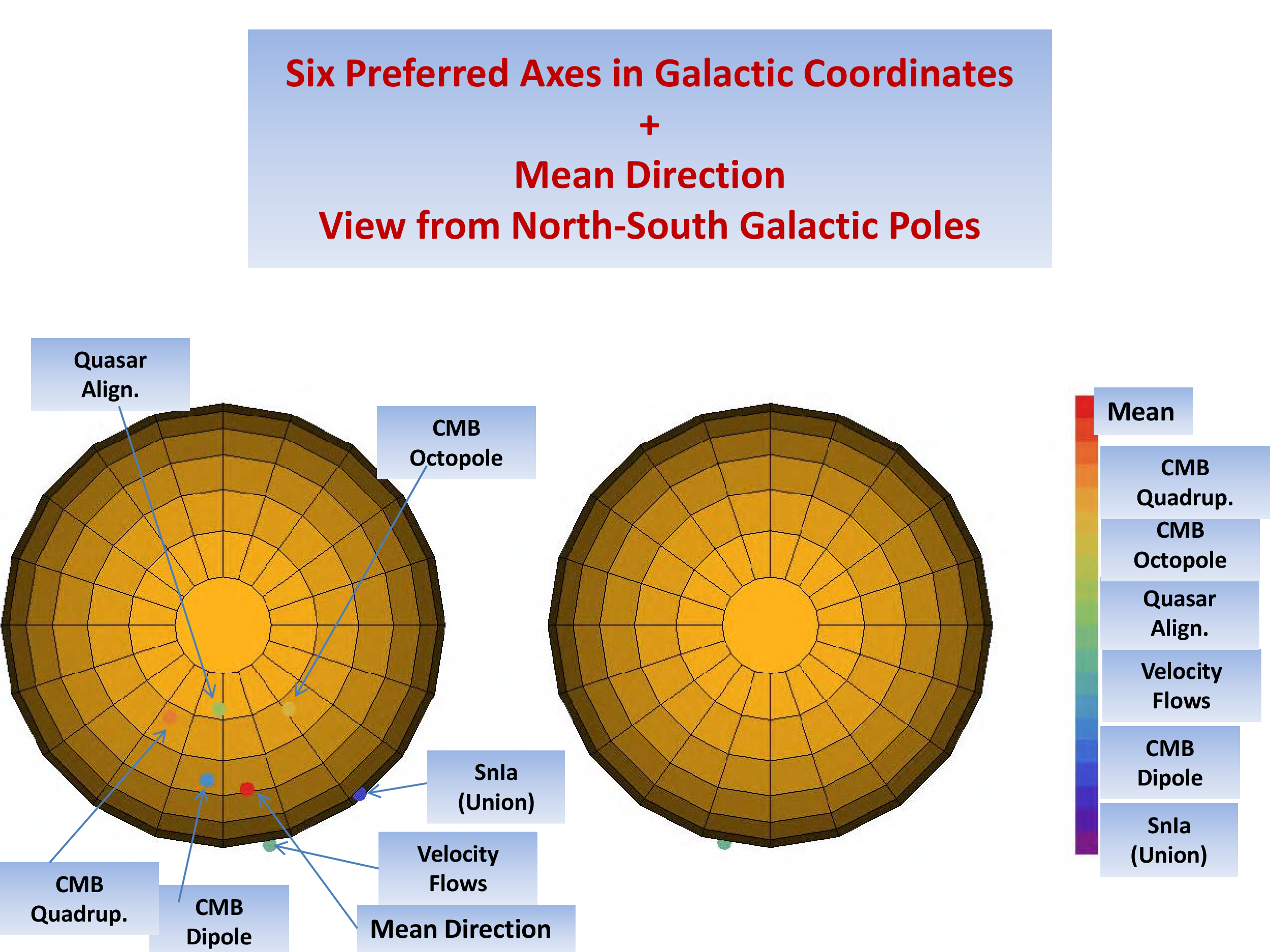}
\end{center}
\caption{\small The coordinates of the preferred axes of Table 1 are all located in a region less than a quarter of the North Galactic Hemisphere (left). The south galactic hemisphere (right) is also shown for completeness. The bulk flow direction is also visible in the south galactic hemisphere because it is close to the equator. The mean direction obtained in Table 1 with coordinates $(l,b)=(278^\circ,45^\circ)$ is also shown.}
  \label{fig4}
\end{figure}
These directions are also shown in Figure 1 in galactic coordinates. It is straightforward to show (Antoniou \& Perivolaropoulos, 2010) that the probability of such proximity among axes directions that should be independent of each other is less than $1\%$.
Thus, unless there is a hidden common systematic (Peiris \& Smith, 2010), the existence of a cosmological preferred axis may be attributed to physical effects. An incomplete list of these effects is the following:
\begin{itemize}
\item
An anisotropic dark energy equation of state (Zumalacarregui {\it et. al.}, 2010; Koivisto \& Mota, 2006; Battye \& Moss, 2009) due perhaps to the existence of vector fields (Armendariz-Picon, 2004; Esposito-Farese, Pitrou \& Uzan, 2010).
\item
Dark Energy and/or Dark matter perturbations on scales comparable to the horizon scale (Rodrigues, 2008; Jimenez  \& Maroto, 2009). For example an off center observer in a 1Gpc void would experience the existence of a preferred cosmological axis through the Lemaitre-Tolman-Bondi metric (Alexander {\it et. al.}, 2009; Garcia-Bellido \& Haugboelle, 2008;  Dunsby {\it et. al.}, 2010; Garfinkle, 2010).
\item
Turbulent structure formation could also lead to large scale non-Gaussian features which would lead to the existence of a preferred axis (Schild \& Gibson, 2008).
\item
Deviations from the isotropic cosmic expansion rate induced by a fundamental violation of the cosmological principle eg through a multiply connected non-trivial cosmic topology (Luminet, 2008), rotating universe coupled to an anisotropic scalar field (Carneiro \&  Mena Marugan, 2001), non-commutative geometry (Akofor {\it et. al.}, 2008) or simply a fundamental anisotropic curvature (Koivisto {\it et. al.}, 2011).
\item
Statistically anisotropic primordial perturbations (Armendariz-Picon, 2007; Pullen \& Kamionkowski, 2007;  Ackerman, Carroll \&  Wise 2007). For example, inflationary perturbations induced by vector fields (Dimopoulos {\it et. al.}, 2009; Bartolo {\it et. al.}, 2009). Note however that inflationary models with vector fields usually suffer from instabilities due to the existence of ghosts (Himmetoglu, Contaldi \& Peloso, 2009).
\item
The existence of a large scale primordial magnetic field (Kahniashvili, Lavrelashvili \& Ratra, 2008; Barrow, Ferreira \& Silk, 1997; Campanelli, 2009). Evidence for such a magnetic field has recently been found in CMB maps (Kim \& Naselsky, 2009).
\end{itemize}
The confirmation of the existence of a cosmological preferred axis would constitute a breakthrough in cosmological research. Given the present status of cosmological observations such a confirmation is one of the most probable directions from which new physics may emerge.

Given the preliminary evidence for anisotropy discussed above, it is important to extend and intensify efforts for the possible confirmation of this evidence. Such confirmation may be achieved by extending the SnIa compilations towards larger datasets and deeper redshifts that span as uniformly as possible all directions in the sky. This is important in view of the fact that the Union2 compilation is less uniform and detailed in the south galactic hemisphere. In addition it is important to extend other cosmological data related to CMB low multipole moments, bulk velocity flows and quasar polarization to confirm the present existing evidence for preferred axes in these datasets. Finally, alternative probes of cosmological anisotropies may be considered like higher CMB multipole moments, non-gaussian features and polarization in the CMB maps, alignments of geometric features of various structures on large scales (there is already some preliminary evidence for alignment of handedness of spiral galaxies (Longo, 2009) along an axis not far from the directions of the other preferred axes of Table 1), alignment of optical polarization from various cosmological sources or studies based on cosmic parallax (Quartin \& Amendola, 2010). It is also important to derive observational signatures that can clearly distinguish between the various different origins of the preferred axes.

{\bf Ackowledgements:} I thank the organizers of the 'New Directions in Modern Cosmology' workshop for inviting me. I have enjoyed many interesting discussions with the participants and the organizers especially with Rick Watkins, Ruth Durrer and Rudy Schild.

\section*{References}
\medskip
\noindent
Ackerman, L., Carroll, S.~M.  and Wise, M.~B. (2007). Imprints of a Primordial Preferred Direction on the Microwave Background. Phys.\ Rev.\  D {\bf 75}, 083502
  [Erratum-ibid.\  D {\bf 80}, 069901 (2009)].

\vskip 2mm
\noindent
Akofor, E., Balachandran, A.~P., Jo, S.~G.,  Joseph, A. and Qureshi, B.~A. (2008). Direction-Dependent CMB Power Spectrum and Statistical Anisotropy from Noncommutative Geometry. JHEP {\bf 0805}, 092.

\vskip 2mm
\noindent
Alexander, S., Biswas, T., Notari A. and Vaid, D. (2009). Local Void vs Dark Energy: Confrontation with WMAP and Type Ia Supernovae. JCAP {\bf 0909}, 025.

\vskip 2mm
\noindent
Allen, S.~W., Schmidt, R.~W., Ebeling, H., Fabian A.~C. and van Speybroeck, L. (2004). Constraints on dark energy from Chandra observations of the largest relaxed galaxy clusters. Mon.\ Not.\ Roy.\ Astron.\ Soc.\  {\bf 353}, 457.
\vskip 2mm
\noindent
Amendola, L., Kunz M. and Sapone D. (2008). Measuring the dark side (with weak lensing). JCAP {\bf 0804}, 013.
\vskip 2mm
\noindent
Antoniou, I. and Perivolaropoulos, L. (2010). Searching for a Cosmological Preferred Axis: Union2 Data Analysis and Comparison with Other Probes. JCAP {\bf 1012}, 012.
\vskip 2mm
\noindent
Armendariz-Picon, C. (2004). Could dark energy be vector-like? JCAP {\bf 0407}, 007.
\vskip 2mm
\noindent
Armendariz-Picon, C. (2007). Creating Statistically Anisotropic and Inhomogeneous Perturbations. JCAP {\bf 0709}, 014.
\vskip 2mm
\noindent
Astier, P. {\it et al.}  [The SNLS Collaboration] (2006). The Supernova Legacy Survey: Measurement of $Omega_M, Omega_Lambda$ and from the First Year Data Set. {\it Astron.\ Astrophys.}  {\bf 447}, 31.
\vskip 2mm
\noindent
Barrow, J.~D., Ferreira P.~G. and Silk, J. (1997). Constraints on a Primordial Magnetic Field. Phys.\ Rev.\ Lett.\  {\bf 78}, 3610.
\vskip 2mm
\noindent
Bartolo, N., Dimastrogiovanni, E., Matarrese S. and Riotto, A. (2009). Anisotropic bispectrum of curvature perturbations from primordial non-Abelian vector fields. JCAP {\bf 0910}, 015.
\vskip 2mm
\noindent
Basilakos S.  and Perivolaropoulos L. (2008). Testing GRBs as Standard Candles. Mon.\ Not.\ Roy.\ Astron.\ Soc.\  {\bf 391}, 411.
\vskip 2mm
\noindent
Battye, R. and Moss, A. (2009). Anisotropic dark energy and CMB anomalies. Phys.\ Rev.\  D {\bf 80}, 023531.
\vskip 2mm
\noindent
Benjamin, J. {\it et al.} (2007). Cosmological Constraints From the 100 Square Degree Weak Lensing Survey. Mon.\ Not.\ Roy.\ Astron.\ Soc.\  {\bf 381}, 702.
\vskip 2mm
\noindent
Bertschinger, E. (2006). On the Growth of Perturbations as a Test of Dark Energy. Astrophys.\ J.\  {\bf 648}, 797.
\vskip 2mm
\noindent
Bielewicz, P., Gorski K.~M.  and Banday, A.~J. (2004). Low order multipole maps of CMB anisotropy derived from WMAP. Mon.\ Not.\ Roy.\ Astron.\ Soc.\  {\bf 355}, 1283.
\vskip 2mm
\noindent
Broadhurst, T.~J., Takada, M., Umetsu, K., Kong, X., Arimoto, N.,  Chiba, M. and Futamase, T. (2005). The Surprisingly Steep Mass Profile of Abell 1689, from a Lensing Analysis of Subaru Images. Astrophys.\ J.\  {\bf 619}, L143.
\vskip 2mm
\noindent
Campanelli, L. (2009). A Model of Universe Anisotropization. Phys.\ Rev.\  D {\bf 80}, 063006.
\vskip 2mm
\noindent
Carneiro S. and Mena Marugan, G.~A. (2001). Anisotropic cosmologies containing isotropic background radiation.  Phys.\ Rev.\  D {\bf 64}, 083502.
\vskip 2mm
\noindent
Copeland E. J., Sami M. and Tsujikawa S. (2006). Dynamics of dark energy. {\it  Int.\ J.\ Mod.\ Phys.\  D} {\bf 15}, 1753.
\vskip 2mm
\noindent
Copi, C. J., Huterer, D., Schwarz, D. J. and Starkman G. D. (2006). On the large-angle anomalies of the microwave sky. Mon.\ Not.\ Roy.\ Astron.\ Soc.\  {\bf 367}, 79.
\vskip 2mm
\noindent
Copi, C., Huterer, D., Schwarz D. and Starkman, G. (2007). The Uncorrelated Universe: Statistical Anisotropy and the Vanishing Angular Correlation Function in WMAP Years 1-3. Phys.\ Rev.\  D {\bf 75}, 023507.
\vskip 2mm
\noindent
Copi, C.~J., Huterer, D., Schwarz D.~J. and Starkman, G.~D.  (2010). Large-angle anomalies in the CMB. Adv.\ Astron.\  {\bf 2010}, 847541.
\vskip 2mm
\noindent
Dimopoulos, K., Karciauskas, M., Lyth D.~H. and Rodriguez, Y. (2009). Statistical anisotropy of the curvature perturbation from vector field perturbations. JCAP {\bf 0905}, 013.
\vskip 2mm
\noindent
Dunsby, P., Goheer, N., Osano, B. and Uzan, J.~P. (2010). How close can an Inhomogeneous Universe mimic the Concordance Model?  JCAP {\bf 1006}, 017.
\vskip 2mm
\noindent
Efstathiou, G., Sutherland, W. J. and Maddox, S. J. (1990). The cosmological constant and cold dark matter. Nature {\bf 348}, 705.
\vskip 2mm
\noindent
Esposito-Farese, G., Pitrou C. and Uzan, J.~P. (2010). Vector theories in cosmology. Phys.\ Rev.\  D {\bf 81}, 063519.
\vskip 2mm
\noindent
Garcia-Bellido J. and Haugboelle, T. (2008). Confronting Lemaitre-Tolman-Bondi models with Observational Cosmology. JCAP {\bf 0804}, 003.
\vskip 2mm
\noindent
Garfinkle, D. (2010).  The motion of galaxy clusters in inhomogeneous cosmologies. Class.\ Quant.\ Grav.\  {\bf 27}  065002.
\vskip 2mm
\noindent
Gentile, G., Salucci, P., Klein, U.,  Vergani, D. and Kalberla, P. (2004). The cored distribution of dark matter in spiral galaxies. Mon.\ Not.\ Roy.\ Astron.\ Soc.\  {\bf 351}, 903.
\vskip 2mm
\noindent
Gorski K. M., Hansen F. K. and Lilje, P.~B. (2007). Hemispherical power asymmetry in the three-year Wilkinson Microwave Anisotropy Probe sky maps. Astrophys.\ J.\  {\bf 660}, L81.
\vskip 2mm
\noindent
Hawkins, E.  {\it et al.} (2003). The 2dF Galaxy Redshift Survey: correlation functions, peculiar velocities and the matter density of the Universe. Mon.\ Not.\ Roy.\ Astron.\ Soc.\  {\bf 346}, 78.
\vskip 2mm
\noindent
Hicken, M. {\it et al.} (2009). Improved Dark Energy Constraints from ~100 New CfA Supernova Type Ia Light Curves. Astrophys.\ J.\  {\bf 700}, 1097.
\vskip 2mm
\noindent
Himmetoglu, B., Contaldi C.~R., and Peloso, M. (2009). Instability of anisotropic cosmological solutions supported by vector fields. Phys.\ Rev.\ Lett.\  {\bf 102}, 111301.
\vskip 2mm
\noindent
Hutsemekers, D., Cabanac, R., Lamy, H. and Sluse, D. (2005). Mapping extreme-scale alignments of quasar polarization vectors. Astron.\ Astrophys.\  {\bf 441}, 915.
\vskip 2mm
\noindent
Jimenez J.~B., and Maroto, A.~L. (2009). Large-scale cosmic flows and moving dark energy. JCAP {\bf 0903}, 015.
\vskip 2mm
\noindent
Kahniashvili, T., Lavrelashvili G. and Ratra, B. (2008). CMB Temperature Anisotropy from Broken Spatial Isotropy due to an Homogeneous Cosmological Magnetic Field. Phys.\ Rev.\  D {\bf 78}, 063012.
\vskip 2mm
\noindent
Kashlinsky A., Atrio-Barandela, F.,  Kocevski, D., and Ebeling H. (2009). A measurement of large-scale peculiar velocities of clusters of galaxies: results and cosmological implications. Astrophys.\ J.\  {\bf 686}, L49.
\vskip 2mm
\noindent
Kim J. and Naselsky, P. (2009). Alfven turbulence in the WMAP 5 year data and a forecast for the PLANCK. JCAP {\bf 0907}, 041.
\vskip 2mm
\noindent
Klypin, A.~A., Kravtsov, A.~V., Valenzuela, O. and Prada, F. (1999). Where are the missing galactic satellites? Astrophys.\ J.\  {\bf 522}, 82.
\vskip 2mm
\noindent
Koivisto T. and Mota, D.~F. (2006). Dark Energy Anisotropic Stress and Large Scale Structure Formation. Phys.\ Rev.\  D {\bf 73}, 083502.
\vskip 2mm
\noindent
Koivisto, T.~S., Mota, D.~F., Quartin M. and Zlosnik, T.~G. (2011). On the Possibility of Anisotropic Curvature in Cosmology. Phys.\ Rev.\  D {\bf 83}, 023509.
\vskip 2mm
\noindent
Komatsu E. {\it et al.} [WMAP Collaboration] (2009). Seven-Year Wilkinson Microwave Anisotropy Probe (WMAP) Observations: Cosmological Interpretation. {\it Ap.J. Suppl.}  {\bf 180}, 330.
\vskip 2mm
\noindent
Kovetz, E.~D.,  Ben-David, A. and Itzhaki, N. (2010). Giant Rings in the CMB Sky. Astrophys.\ J.\  {\bf 724}, 374.
\vskip 2mm
\noindent
Kowalski, M. {\it et al.} (2008).  Improved Cosmological Constraints from New, Old and Combined Supernova Datasets. Astrophys.\ J.\  {\bf 686}, 749.
\vskip 2mm
\noindent
Krauss, L. M. and Chaboyer, B. (2003). Age Estimates of Globular Clusters in the Milky Way: Constraints on Cosmology. Science {\bf 299}, 65.
\vskip 2mm
\noindent
Land, K. and Magueijo, J. (2005). The axis of evil. Phys.\ Rev.\ Lett.\  {\bf 95}, 071301 (2005).
\vskip 2mm
\noindent
Lavaux, G.  Tully, R. B., Mohayaee, R. and Colombi, S. (2010). Cosmic flow from 2MASS redshift survey: The origin of CMB dipole and implications for LCDM cosmology. Astrophys.\ J.\  {\bf 709}, 483.
\vskip 2mm
\noindent
Lineweaver, C.~H., Tenorio, L., Smoot, G.~F., Keegstra, P., Banday, A.~J. and Lubin, P. (1996). The Dipole Observed in the COBE DMR Four-Year Data. Astrophys.\ J.\  {\bf 470}, 38.
\vskip 2mm
\noindent
Longo, M.~J. (2009).  Evidence for a Preferred Handedness of Spiral Galaxies. arXiv:0904.2529.
\vskip 2mm
\noindent
Luminet, J.~P. (2008).  The Shape and Topology of the Universe. arXiv:0802.2236 [astro-ph].
\vskip 2mm
\noindent
McDonald, P. {\it et al.}  [SDSS Collaboration] (2005). The Linear Theory Power Spectrum from the Lyman-alpha Forest in the Sloan Digital Sky Survey.  Astrophys.\ J.\  {\bf 635}, 761.
\vskip 2mm
\noindent
Nesseris, S. and Perivolaropoulos, L. (2008). Testing LCDM with the Growth Function $\delta(a)$: Current Constraints. Phys.\ Rev.\  D {\bf 77}, 023504.
\vskip 2mm
\noindent
Padmanabhan, T (2003). Cosmological constant: The weight of the vacuum. Phys.\ Rept.\  {\bf 380}, 235.
\vskip 2mm
\noindent
Peiris H.~V. and Smith, T.~L. (2010). CMB Isotropy Anomalies and the Local Kinetic Sunyaev-Zel'dovich Effect. Phys.\ Rev.\  D {\bf 81}, 123517.
\vskip 2mm
\noindent
Perivolaropoulos, L. (2008). Six Puzzles for LCDM Cosmology.  Invited article to the TSPU anniversary volume "The Problems of Modern Cosmology" on the occasion of the 50th birthday of Prof. S. D. Odintsov. arXiv:0811.4684.
\vskip 2mm
\noindent
Perlmutter S. {\it et al.}  [Supernova Cosmology Project Collaboration] (1999). Measurements of Omega and Lambda from 42 High-Redshift Supernovae. Astrophys.\ J.\  {\bf 517}, 565.
\vskip 2mm
\noindent
Plionis, M., Terlevich, R.,  Basilakos, S., Bresolin, F., Terlevich, E., Melnick, J. and Georgantopoulos, I. (2009). Alternative High-z Cosmic Tracers and the Dark Energy Equation of State. J.\ Phys.\ Conf.\ Ser.\  {\bf 189}, 012032.
\vskip 2mm
\noindent
Pullen A.~R. and Kamionkowski, M.  (2007). Cosmic Microwave Background Statistics for a Direction-Dependent Primordial Power Spectrum. Phys.\ Rev.\  D {\bf 76}, 103529.
\vskip 2mm
\noindent
Quartin M. and Amendola, L. (2010). Distinguishing Between Void Models and Dark Energy with Cosmic Parallax and Redshift Drift. Phys.\ Rev.\  D {\bf 81}, 043522.
\vskip 2mm
\noindent
Rapetti, D., Allen, S.~W., Mantz A. and Ebeling, H. (2008). Constraints on modified gravity from the observed X-ray luminosity function of galaxy clusters. arXiv:0812.2259 [astro-ph].
\vskip 2mm
\noindent
Reid B.~A. {\it et al.}  [SDSS Collaboration] (2010). Baryon Acoustic Oscillations in the Sloan Digital Sky Survey Data Release 7 Galaxy Sample. Mon.\ Not.\ Roy.\ Astron.\ Soc.\  {\bf 401}, 2148.
\vskip 2mm
\noindent
Rodrigues, D.~C. (2008). Anisotropic Cosmological Constant and the CMB Quadrupole Anomaly. Phys.\ Rev.\  D {\bf 77}, 023534.
\vskip 2mm
\noindent
Sahni, V. (2002). The cosmological constant problem and quintessence.Class.\ Quant.\ Grav.\  {\bf 19}, 3435.
\vskip 2mm
\noindent
Schild R.~E. and Gibson C.~H. (2008). Goodness in the Axis of Evil. arXiv:0802.3229 [astro-ph].
\vskip 2mm
\noindent
Tegmark, M., de Oliveira-Costa, A. and Hamilton A. (2003). A high resolution foreground cleaned CMB map from WMAP. Phys.\ Rev.\  D {\bf 68}, 123523.
\vskip 2mm
\noindent
Tsagas, C. G. (2010). Large-scale peculiar motions and cosmic acceleration. Mon.\ Not.\ Roy.\ Astron.\ Soc.\  {\bf 405}, 503.
\vskip 2mm
\noindent
Umetsu, K. and Broadhurst, T. (2008). Combining Lens Distortion and Depletion to Map the Mass Distribution of A1689. Astrophys.\ J.\  {\bf 684}, 177.
\vskip 2mm
\noindent
Watkins, R., Feldman, H. A. and Hudson, M. J. (2009). Consistently Large Cosmic Flows on Scales of 100 Mpc/h: a Challenge for the Standard LCDM Cosmology.  Mon.\ Not.\ Roy.\ Astron.\ Soc.\  {\bf 392}, 743.
\vskip 2mm
\noindent
Zumalacarregui, M., Koivisto, T.~S., Mota, D.~F. and Ruiz-Lapuente, P. (2010). Disformal Scalar Fields and the Dark Sector of the Universe. JCAP {\bf 1005}, 038.


\end{document}